# Your Interests According to Google – A Profile-Centered Analysis for Obfuscation of Online Tracking Profiles

Martin Degeling

Ruhr-Universität Bochum

martin.degeling@ruhr-uni-bochum.de

Thomas Herrmann

Ruhr-Universität Bochum

thomas.herrmann@iaw.ruhr-uni-bochum.de

*Abstract* – **Profiling users for the purpose of targeted advertisements or other kinds of personalization is very popular on the internet. But besides the benefits of individually tailored news feeds and shopping recommendation research has shown that many users consider this practice as privacy infringement. Profiling is often conducted without consent and services offer neither information about what the profile looks like nor which effects it might have. In this paper we argue that understanding profiling and thus interacting with the resulting profiles fosters a privacy literacy that is necessary for users to stay autonomous in the information society. We analyze the extent of interest profiling by google and develop a counter-measures that helps to obfuscate these profiles. To do so we analyzed a links lists from a social bookmarking service with regard to the interests they reveal. We found that, although the profiling by google is very unstable we can still use the information to obfuscate the profile.**

## I. Introduction

Personalized recommendation and tailoring of information has become a widespread functionality to make services more effective and efficient. They are adopted by a broad range of online services from social networks to shopping websites. However, to achieve better personalization the services process more and more data about users who are often not aware of this. If this *profiling* takes place without explicit consent and in ways that are not transparent to users it is a threat to (informational) privacy. A recent study also showed that 55% of US internet users don't think that there is a fair trade-off between benefits and their privacy, although they regularly give their consent to this practices [1]. In most cases they do not have a sufficient understanding of what the assumed characteristics of their digital identities – also known as *data doubles* [2] - are, nor are there tools that effectively support this understanding [3]. While some see benefits in personalized recommendations most users just dislike and ignore common forms of *online behavioral advertisement (OBA)* [4], [5]. Several researches have addressed the negative side effects of profiling like discrimination [6], [7] or filter bubbles [8]. On a more general level the effects of tracking can also be described as a threat to the autonomy of internet users as privacy can be regarded as a mechanism to protect individuals' autonomy and control when using technical systems [9]. Tracking technology is not only used for delivering personalized ads but also to alter the content of websites [10]. Some even assume that it is used to alter prices to motivate customers in buying a product [11], [12]. This is on the one hand thought to anticipate the interests of users but on the other hand also fits the needs of service providers to increase sales or reduce risks. However, this can, in the end, result in an exclusion of participation e.g. when products are simply not offered to those that live in a specific region [13]. With the rising use of ad blockers now residing between 15 and 35% of all internet users [14] services are starting to prevent users from using this effective anti-profiling technique. They start to block ad blocker users [15] or circumvent them with new tracking techniques. In addition, blockers often reduce the functionality of a website and do not offer any means to understand the profiles that are built and the effects they have.

A strategy to actively influence online tracking is known as obfuscation [16]. In contrast to blocking specific parts of websites obfuscation of web trails tries to hide the real interests of a user in dummy traffic and therefore obfuscates her profile. Obfuscation is especially useful when blocking of websites or parts of websites would limit the usefulness of a platform. So far obfuscation techniques were developed for specific settings like web search [17], [18], location based-services [19], recommender systems [20] and online social networks [21]. We extend this research by gaining knowledge about the profiles that are created by online trackers focusing not on the tracking itself but on common surfing behavior. We want to use this knowledge to empower users to control these profiles. More specifically, the contribution of this paper is

i.  an **analysis of tracking and interest profiling** performed by Google services on users who are not logged in **based on authentic browsing histories** and

ii. using this information to create an **informed Dummy Generation Strategy** (DGS) [22] for profile obfuscation that influences profiles based on web navigation trails.

In addition, we want to address the critique that the effect of theoretic obfuscation models is often not measurable when applied in a real world environment [16]. We demonstrate how the effect on the breadth of profiles created by Google can be evaluated.

To measure effects of our strategies we created an automatic browsing process used to analyze 500 URL lists publicly posted by Reddit users. We made the extent and impact of interest profiling by Google visible. Google's servers follow users on 81% of their web history and extract 17 interests on average. The information provided by this study can be used to give feedback to users on how their interaction with an online service will influence the future of the user's and others' interactions with this service. Additionally, we also demonstrate that and how this influence can be controlled by the users through obfuscation

The rest of the paper is organized as follows: first we will describe some perspectives on the discussion about profiling and the negative effects on users. Then we will present how we analyzed the extent of profiling by Google and the interest profiles it generates. The paper concludes with an evaluation of obfuscation based on the learned connection between interests and specific web sites.

## II. Background

One of the tracking providers, Acxiom, describes in a presentation how it not only targets users with personalized services, but also tries to influence them by offering discounts [23]. Advertising in this case takes it a step further, from trying to deliver (visual) information towards triggering a specific behavior by making assumptions about interests or shopping behavior based on a rather small set of information that is extracted from the browser while surfing the web. Similarly, a Facebook employee said regarding their reasons to build user profiles that they "[...] try to understand what kind of characteristics make someone likely to take a certain type of action. When [an advertiser] specifies a desired action – say registration or purchase – we look for users more likely to take that action."[24] These assumptions may for example be that persons categorized as young and mobile males respond more often to discounts while those categorized as female and older do more often buy products that are offered with free shipping.

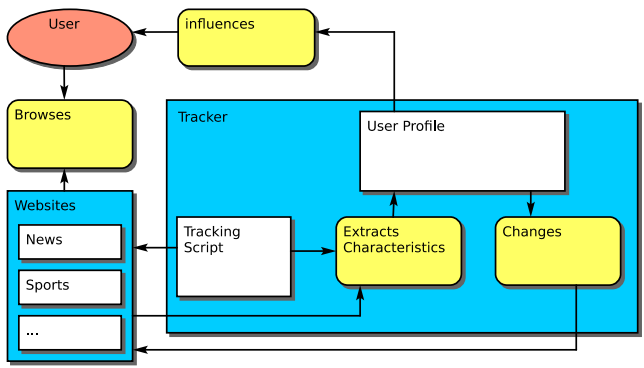

*Figure 1: Abstract model of the feedback loop of online tracking and profiling*

In most cases this is only a problem when it comes to consuming and online shopping, but similar mechanisms are in place when information on news sites are tailored to assumed interests. Gillespie [25] argued that this effect, later coined the *filter bubble* [10], may lead to less comprehensive understanding of events, since one cannot decide which information source to trust when they are preselected by an algorithm. Automatic assumptions about characteristics are used to change which information is delivered and what options can be chosen is a threat to what Rössler named as *decisional privacy* [9].

Galloway [26] describes a feature of these mechanisms as *cycles of anticipation* (see Figure 1[1]) since they do not operate on static data that is collected from the user once e.g. when profile information is entered voluntarily during registration. Instead, it relies on dynamic data that is continuously created by tracking millions of users. This information is then used to try to influence the individual and again collecting her behavior traces. In general, the feedback loop within online profiling does not care for the identity of the user. User characteristics are estimated based on traces like IP-based geo-location, browser settings and surf history. They are then enriched with data from other sources to which a user may never have contributed to individually. In these cases, the predicted behavior, especially if a user was influenced successfully, is more relevant than the actual characteristics of the user in the real world. In a *feedback loop* each action taken is itself an information that feeds back into the system which is used to choose the possible actions that are available in the next step. In this way of thinking the *cybernetic hypothesis* [27] every user is regarded as a black-box that can be regulated by the input (e.g. ads or discounts) and not as an autonomous person, for which the output has to be brought in line with the trackers' goal (e.g. conversion rate). In this system, every action of a user or of the responding algorithms provide feedback to each other [28]. The problem is that there is no direct option for a user to determine which of her actions – and to what extent – produce a specific reaction by the advertising delivery algorithm, whilst the algorithm is open to be influenced by its designers, e.g. to favor specific content and push a user in a specific direction.

Although the problematic aspects of profiling have been discussed [29], [30] and users widely dislike these practices [4], [5], there is a lack of options for users in their daily action. Especially on the policy level there is currently no possibility to prevent profiling if the data used is anonymous or aggregated. Initiatives such as the DoNotTrack (DNT) HTTP-header lack wide spread adoption [31] and even are getting rejected by online advertisers. Opt-out options provided by advertising networks like AdChoices[2] only offer the possibility to end the personalization of ads but won't stop the tracking and profiling itself. From a legal perspective in most cases tracking users anonymously is considered to be in compliance with international law as long as a re-identification is not possible.

### A. Related Work on Online Tracking

Online tracking occurs in various forms from which tracking with browser cookies is still the most common technology [32]. Besides simple implementation its outreach increases these days through the use of Social media plugins and Content Delivery Networks (CDN) that are included as third-party elements in many websites. In these cases, functionality on websites requires that scripts and/or images are requested from third party web servers which comes along with the transfer of a client-identifying (not person identifying) text string called cookie. However, because not all cookies are used for tracking purposes, they cannot be totally blocked. Cookies are important for managing sessions for users that logged into a server or to store preferences (e.g. to select a non-default language). Countermeasures as to block third-party traffic and cookies is therefore difficult to implement without accidentally blocking essential parts of the web experience. In addition, other forms of third-party tracking are emerging that are ever more difficult to prevent (for example tracking with *etags* [33], various forms of *zombie cookies* [34] or combination of various Browser information known as *fingerprinting* [35]).

The latter, believed to be a more theoretical possibility, was recently studied and found to be used 'in the wild' [36]. A combination of multiple tracking technologies is also used to track even those users who try to opt-out by deleting their cookies, last shown by Acar et.al [37]. Methods other than tracking cookies are on the rise and not only much more difficult to detect, but also more difficult to avoid since they do not rely on information that is stored on a user's computer like cookies, but are inherent in browsers and HTTP itself [38]. Considering the Snowden revelations, blocking online trackers does not prevent those capable of accessing the whole traffic bound to an IP address from compiling it into profile. Therefore, a reasonable response may be to not try to limit the amount of data available but to increase it, to hide within the noise and obfuscate ones traces. While there is considerable research on technical details of tracking and blocking, questions remain on how successful tracking and profiling works. Especially information on how profiles are created and which data is used to build them can be used to foster better obfuscation techniques.

---

[1] The model is based on the semi-structured modeling language SeeMe. Round elements represent roles, those with round edges activities and those with sharp edges resources. Arrow indicate relations and influences.

[2] See http://www.youradchoices.com/ which refers to http://www.aboutads.info/choices/ to opt-out of 120 ad services.

There are first steps of analyzing the information flow for online advertisements [39] and the relation between Gmail and ads [40] as well as transparency enhancements in the browser itself [41]. The authors point out the effects of profiles created for behavioral targeting and contribute to a better understanding of how profiling works, but they do not use this information to influence the profiles. In this paper we will first elaborate on the interest profiles used by Google as one of the major participants in this market. And second, we want to address the question: How can insights in the actual practice of profiling be used to intervene in these processes in a way that can offer users transparency about and control over the profiles that are created?

*B. Related Work on Obfuscation*

Obfuscation is often referred to as a privacy enhancing technology (PET) "which offers a strategy for mitigating the impact of the cycle of monitoring, aggregation, analysis, and profiling, adding noise to an existing collection of data in order to make the collection more ambiguous, confusing, harder to use, and therefore less valuable"[16]. With obfuscation profiles are similar to masks which allow users to use the internet without being identified. Unlike those users that block profiling and are therefore identifiable as those who are without a profile[3]. With the ability to change profiles like masks, a user is enabled to interact with a profiling based service as if she was somebody else. Therefore, she avoids the filter bubble by using multiple filters. By broadening the profile, the user increases the possibility of serendipity effects during her web research.

A widely discussed example for those PET is TrackMeNot [17], [18], a Firefox plugin that can be used to obfuscate search queries on Google. TrackMeNot continuously sends random or dummy queries to Google to hide real queries made by a user within this noise. In the continuing work the authors focused on enhancing the dummy generation strategy to be less obvious so that a search provider cannot simply filter out dummy traffic. Users are also enabled to narrow the topics about which dummy traffic is generated to avoid issuing queries that might not be acceptable in some circumstance (e.g. at work). Nevertheless, obfuscation tools lack a wide adoption. They require a user to already know about profiling and offer no means to see what the obfuscated profiles actually look like.

Besides proposing ways for obfuscation of online tracking as a way to strengthening informational privacy by fostering users' abilities to decide on how they want to be recognized, we also want to address previously identified challenges [16]. The authors ask, whether it is possible to quantify and optimize different tactics of obfuscation. The approach presented in our paper takes some steps in this direction by introducing a mean to measure and improve obfuscation effectiveness based on information provided by the trackers. This approach can be used for different tactics of obfuscation such as 'hiding within noise' to be unrecognizable and 'masking or impersonating' with a different profile that is in itself valid. We do so, not by reverse engineering profile generation, but rather by looking at the outcomes of tracking, in particular in how the user's input (the web navigation trail) is transformed into an interest profile.

Balsa et al. [22] developed a model to analyze web search obfuscation tools. As described above, one key element of obfuscation is the "dummy generation strategy" (DGS) that generates dummy web traffic – in the case of web search it generates dummy search requests to hide regular user requests within a large amount of automatically randomized requests. However, an attacker who wants to filter out the dummy traffic to build up a profile based on the real requests could do this with a profile filtering algorithm (PFA). According to Balsa et al. this filtering may target two different vulnerabilities. First a *profile based analysis* could identify dummy traffic by comparing a profile that includes dummy traffic with regular profiles known to be not obfuscated. Dummy traffic that is too broad in a sense that it does not reflect a "normal" user behavior e.g. including less frequent search terms more than others, might therefore be identified. A second strategy referred to as *query based analysis* may identify dummy traffic by elements of the query itself. For example a query that includes meta-data that identifies it as automatically generated can be easily filtered out. For the presented study we neither tried to hide from PFAs nor did we find any evidence that Google has implemented any.

III. STUDYING INTEREST PROFILING BY GOOGLE

The following study is focused on the analysis of the profiles created by Google for advertising purposes. Google advertising and tracking services are widely used in practice and Google itself offered basic options to control and review the profile it creates (see Figure 2) for users that is not logged in to their services and may therefore think she is anonymous. [4] To be able to compare the extent of tracking and profiling by Google on a larger scale we set up an automatic crawling system called SYSTEM-A (see Figure 3Figure 3). We used SYSTEM-A to first collect profiles from Reddit that offer a user created list of URLs and we then used these URLs in an automated crawling process to collect information about the amount of tracking and the resulting profiles generated when accessing the sites. Our main datasets were the user-URL-lists from Reddit, the interest lists from Google and network logs from the automatic browsing.

---

[3] Which may lead to further assumptions. e.g. when a tracker recognizes that a specific browser plugin like ghostery is installed, the user can be categorized as privacy aware

[4] Since July 2015 this services is no longer available for users that have not registered with google although the profiling is still in place.

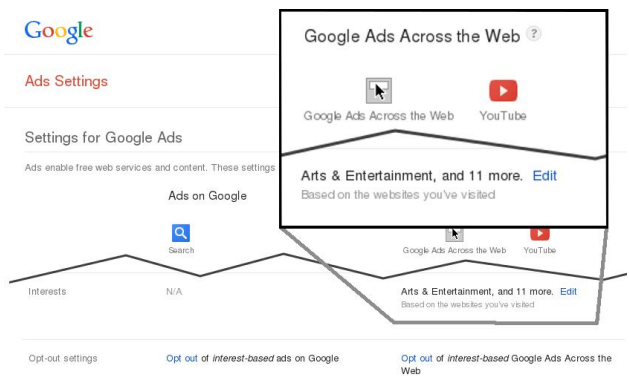

*Figure 2: Googles Ad Settings page showing the interests assigned based on user tracking. Available at http://www.google.com/settings/ads*
*(The page was redesigned July 2015 and is no longer available for users that are not logged)*

### A. Reddit User Data

Since we do not have access to a larger number of browsing histories we turned to Reddit. Reddit is an online community that is used to share and discuss links posted by users. Lists of links are organized in subreddits related to a specific topic where users can vote and comment on them. In addition to these community-based actions, there are also a number of users who uses the site mainly as an individual bookmarking service, posting websites they visited unattached to the idea of reaching a broader audience. We used the bookmarks of those users as input since we assume that they visited these sites at least once and that this link list reflects at least a part of the users' interests when being online. We are aware of the fact, that the user base of Reddit does not reflect the general population, since it is known to be a relatively young, white, male and English speaking community. Nevertheless, the generalizability of our user base is not required since we did not want to study average profiles, but interest profiles that can be linked to a single user, regardless of the socio-demographic attributes of the users.

To focus on the bookmarking users we selected a subsample of 506 users that fulfilled several requirements:

- Included users who posted at least 60-100 links (to exclude inactive users). 100 is the maximum number of links provided per user.
- These 60+ links should point to at least 60 external sites (to exclude users who are heavily engaged in Reddit-internal linking and to exclude those who only use it to be up to date with Internet Memes).
- The links should be diverse in the sense that they should point to at least 20 different domains.

We accessed Reddit through its public API which allows automatic analysis. We used the "random post" API to select the users and analyzed whether they meet the above listed requirements. More than 10.000 profiles had to be processed to get the sample of significant size to serve as an input to the crawling process of accessing all linked pages.

---

[5] Accessible on https://support.google.com/ads/answer/2842480 (last visited 10.11.2015)

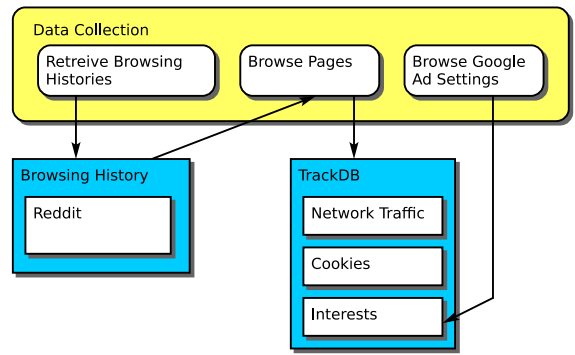

*Figure 3: Basic Elements of the Crawling System SYSTEM-A*

### B. Google Interests Profiles

The Google Interest Profile was extracted from the ad-preferences page (see Figure 2) after a browsing session was completed. On this page Google offered users the possibility to check what the algorithms think they are interested in. It was intended to be used by those that own a Google profile and have manually added information to that profile like gender, age and could be used also by those users that are not logged in, to check which interests are assigned to them by Google's algorithms analyzing the searches they did and websites the visited.

The full list[5] of interest are organized in a tree-structure with multiple categories. In total the list contains 2042 items from which 847 are in the category "World Locations" that is not displayed on the Google ad settings page. The other 1195 interests are split into 24 "root" categories with a varying number (up to 6) sub-categories such as "People & Society/ Family & Relationships/ Family/ Parenting/ Babies & Toddlers/ Baby Care & Hygiene". We called the root categories Google interest categories (GIC). The number of subcategories are listed in Table 1 (left column). While some have about 150 Subcategories (e.g. Arts & Entertainment) others only spread over nine sub-interests (e.g. "Real Estate" or "References")

Based on the interests listed on the Google ad settings page we defined an interest profile for user (U) as the tuple $IP_U = (x_1, \ldots, x_n)$ where $x_i$ is 1 if any sub category of interest category GIC $i$ was listed. And 0 if no sub-category of interest category $i$ was listed.

Therefore, we define the *breadth* of a user's profile as the sum of this tuple.

(1) $$B_U = \sum IP_U$$

### C. Browsing Data

For each user we set up an artificial browsing session. All data collection and analysis was carried out using standard tools. Multiple Debian/Linux based servers hosted a MongoDB Instance and various scripts. Browsing behavior was simulated with the headless browser PhantomJS[6] which surfed the roughly 100 pages and waited 1 minute for each page to render all scripts before the next page was requested. All HTTP Requests were logged together with corresponding cookies and header information. Afterwards the profile folder was moved and surfing the next users' links started on a clean profile.

*Table 1: Distribution of Google Interest Categories (GIC)*

| Interest Category (No. of Subcategories) | Occurrence of any sub-category in that tree for all users |
|---|---|
| Arts & Entertainment (147) | 80% (318) |
| News (21) | 67% (265) |
| Games (42) | 54% (213) |
| Law & Government (36) | 47% (186) |
| Finance (50) | 45% (181) |
| Computers & Electronics (128) | 43% (175) |
| Internet & Telecom (34) | 43% (172) |
| Sports (69) | 42% (171) |
| Business & Industrial (121) | 40% (160) |
| People & Society (40) | 32% (129) |
| Science (25) | 26% (105) |
| Shopping (71) | 24% (96) |
| Travel (27) | 22% (88) |
| Autos & Vehicles (95) | 22% (87) |
| Food & Drink (73) | 21% (83) |
| Beauty & Fitness (21) | 20% (82) |
| Jobs & Education (36) | 20% (80) |
| Reference (30) | 18% (71) |
| Online Communities (18) | 18% (71) |
| Pets & Animals (15) | 16% (65) |
| Books & Literature (9) | 12% (48) |
| Home & Garden (48) | 12% (46) |
| Hobbies & Leisure (30) | 6% (23) |
| Real Estate (9) | 2% (7) |

---

[6] We used PhantomJS version 1.9 that was modified to allow execution of Flash videos. It was also configured to accept all cookies and certificates.

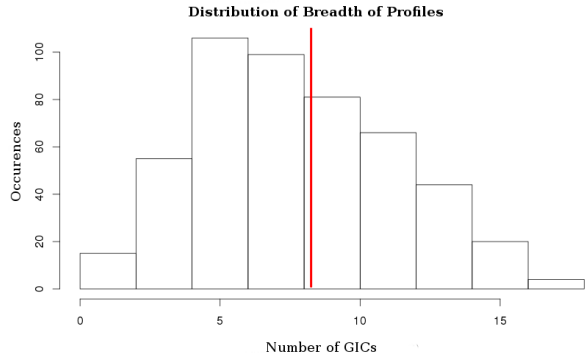

*Figure 4: Distribution of Profile Breadth. Red line indicates*

## IV. TRACKING THE TRACKERS

In the following section we present some descriptive statistics about the user base, the amount of tracking and the profiles that were generated using SYSTEM-A. Afterwards we will use this information as a test-set for SYSTEM-B to evaluate obfuscation mechanisms.

### A. User Statistics

We analyzed the lists of URLs for 506 different users. These lists consisted of 96 URLs in average with a standard deviation (s) of 13.72. Browsing habits of the users are reflected in the fact that these URLs in average linked to 44 different domains (s=14.85). The overall dataset contains 45829 links to 7123 different domains. As could be expected, the number of links to each domain highly varied. The top 100 URLs made up about 49% of all links. Due to the nature of Reddit, that is linking to public content not to content in closed services, in our data sets the most linked web sites after *youtube.com* and *imgur.com* were sites covering news such as *guardian.com, nytimes.com, bbc.co.uk, huffingtonpost.com* and others. The specifics of the links at Reddit is especially striking when compared to the widely recognized Alexa[7] Top 500 which lists the most visited web sites based on the traffic they produce (for a comparison see Table 2).This list is led by search pages (*Google*, *yahoo*, *baidu* or *live.com*) and social networking sites (*facebook*, *linkedin* and *twitter*).

For the purpose of our evaluation the Reddit data set is beneficial since news sites often include third party advertisements (and tracking) scripts which track a user on multiple sites while for example Facebook maintains its own user tracking and advertisement service without directly involving third parties. In addition, third party advertisement services rely on pseudonymous IDs to identify the same user on different web sites and build profiles based on the browsing behavior, while services that require a login, such as Facebook, can enrich the browsing profiles with information entered by the users themselves.

---

[7] Alexa is a Webanalytics company that regularly publishes a widely recognized lists with its own estimates about the most visited sites online. The comparison is with the Top-List from 28.08.2014- See http://www.alexa.com/topsites

## B. Tracking Statistics

PhantomJS allowed us to log all requests the browser issued. A request may be an HTML-page as well as an image or a JavaScript. Nowadays JavaScript-Code issues requests to dynamically load content into web pages. Looking at the requests is of interest for our study because cookies and referrer URLs are part of a standard HTTP request and are used to identify a user. For example, if a user visits *reuters.com* and *huffpost.com* which both implemented Google services, for each visit a JavaScript-file is requested from a Google server. These requests come along with a cookie and a referrer that the Google server can use to make a connection between the user id and the two websites. Based on the Google search index that analyses the content of web pages, the tracker is capable of connecting a topic to the user ID, too. On average 20% of all content that constitutes a web page, are requested from external servers. This includes advertisements but also images, Facebook's "Like" or Twitter's "Tweet"-Buttons, links to Google's Translation Service or trusted-shop logos. In our dataset a median of 641 different cookies was issued to each user. As described above not all cookies are used for tracking purposes.

When it comes to Google, we identified 15 top level domains that can be associated with the company, four are directly related to advertisement and user tracking (*google-analytics.com, doubleclick.[com/net], googlesyndication.com, googleadservices.com*) others offer end-user services (*google.com, youtube.com, youtu.be*) or provide script and font files through a Content Delivery Network (*googleapis.com)*. The dedicated tracking and audience analytic domains are contacted on 75% of an average user's links. When looking at all the 15 domains, the number increases to about 83% of all page visits that were recognized by Google. These results are similar to prior research [42] that analyzed each website independently and found Google services to be contacted on 88% of the websites. This also shows that the Google interest profiles are an appropriate start for analyzing the effectiveness of countermeasures. Other third parties that are embedded in a large percentage (30-50%) of the URLs of each user's list were *facebook.com, twitter.com and scorecardresearch.com* and *quantserver.com*.

## C. Interest Distribution

After the browsing process, we crawled the interests assigned by Google to each user from the ad preferences page. On average each user was assigned 16.34 (s=7.5) interests ranging from 1 to 37 from the 1195 interests available. To be able to better compare the users with each other we looked at the 'google interest root categories' (GICs) instead of the interest leaves as described above. Users' interest stem from 8.07 (s=3.41) of the 24 GIC (see Figure 4). Table 1, right column, shows the percentage of occurrences of a GIC within the dataset. The distribution is largely comparable to the interests one would expect given that the most viewed sites on the web are entertainment or news related as listed e.g. on the Alexa Toplist. However other high ranked interests like "Games" are presumably biased by the sample population.

## D. Limitations

Online advertisements are not only personalized, but the ad space is sold and purchased in a dynamic market referred to as *programmatic advertisement* combined with a strategy called *real-time bidding* [43]. This inner-advertisement-economy is used to purchase advertisement space on various websites leading to websites not knowing which ad publisher will show an ad. For our analysis this means that every time a website is visited, another tracker might be present and while google might serve the ad on one page load, might not do so on the next. It is further possible that Google assigns interests not based on the content of the site but rather by using a relation to the interests that are known about other users of the same site. For example, if user *A* reads a news article about topic 1 that was also read by users *B* and *C*. If users *B* and *C* both own a Google Account which they used in some way to express their interest in topic 2, this might lead to topic 2 (and the corresponding interest) being assigned to user *A* as well. In addition websites that update their content frequently, like those of large news agencies which are prominent in the dataset (*theguardian, nytimes* etc.) lead to a large variety of assigned interests. When we evaluated this variance we even recognized that even for pages that did not change, the assigned interests were not constant. For example, multiple parallel issued requests to the front page of *wired.com* resulted in the interest "Computer and Electronics", for all sessions, but also in up to two other varying interests. Furthermore, [39] proofed that Google is not reporting all information that is used in a profile on the ad settings page.

*Table 2: Top linked sites and the corresponding Alexa Rank*

| No | Domain | # Links | Alexa Rank |
|----|--------|---------|------------|
| 1 | imgur.com | 3173 | 49 |
| 2 | youtube.com | 2725 | 3 |
| 3 | theguardian.com | 1033 | 134 |
| 4 | nytimes.com | 854 | 115 |
| 5 | reuters.com | 686 | 297 |
| 6 | bbc.co.uk | 659 | 62 |
| 7 | washingtonpost.com | 587 | 289 |
| 8 | huffingtonpost.com | 554 | 68 |
| 9 | en.wikipedia.org | 480 | 6 |
| 10 | news.yahoo.com | 376 | 4 |
| 11 | flickr.com | 372 | 107 |
| 12 | reddit.com | 372 | 50 |

## E. Summary of Tracking Analysis

In this section we showed how SYSTEM-A can be used to analyze the extent of tracking by Google services with a focus on the interest profiles that are created. Google is able to create a profile consisting of 7 basic interests out of 75 percent of websites that contact Google's server, mostly without knowledge of the user. Nevertheless, there is no predetermined subset of websites that lead to a specific interest assignment by Google. This is related both to the fact that advertisement spaces are not always filled with ads from the same provider, and that external factors may influence which interests are assigned (e.g. the order websites are surfed). We will therefore use a weighted model to connect interests to websites which can serve two purposes. On the one hand it can be used to foster transparency about the interests a user might get assigned without having to rely on the availability of Google's Ad Settings page on the other hand it serves as input for our obfuscation mechanism described below.

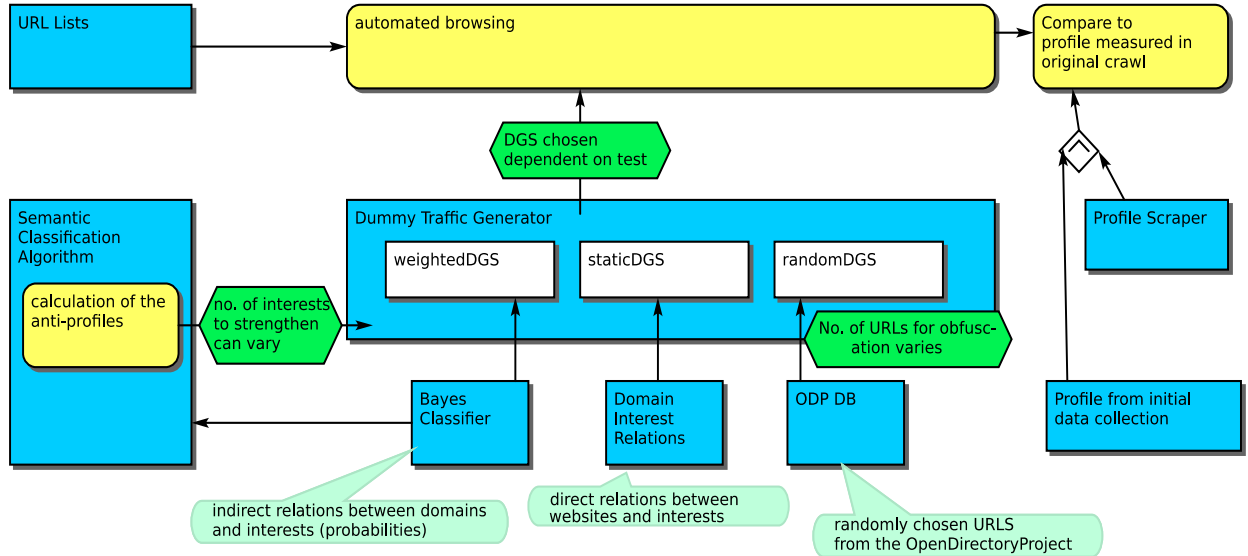

*Figure 5: System Design for Obfuscation (SYSTEM-B)*

## V. OBFUSCATION SYSTEM DESIGN

For now, we learned how profiles look like which can help users to understand why a system responds the way it does. In a second step use the collected data to make an attempt to mitigate the profiling through obfuscation. For SYSTEM-B three dummy generating strategies were evaluated primarily to test and to compare how well they can obfuscate a GIC-Profile.

While the first strategy (randomDGS) was designed for the purpose of comparison the latter two strategies (static and weightedDGS) took into account what we learned from SYSTEM-A.

### A. (Re-)constructing profiles and anti-profiles

To obfuscate a profile we first created a 'real' profile to be able to obfuscate a user regardless of whether or not the profile created by google is known. To do this the list of URLs a user has visited was evaluated and based on the existing data an interest profile was estimated to know which interests are least likely to be part of the profile. This *anti-profile* than served as a basis for the dummy generation strategies. For the construction of the 'real' profile we used the collected data described above to calculate the probabilities of a domain leading to a specific interest. We then checked if these probabilities can be used to reconstruct the profile for a unknown list of URLs. More specifically we:

1. Calculate the probability P for domain $D_i$ supporting on of the interests categories $GIC_j$ by dividing the number $N_{i,j}$ of all co-occurrences of Domain $D_i$ and $GIC_j$ by the number of occurrences of $GIC_j$ with any Domain

$$(1)\ P(D_i|GIC_j) = \frac{N_{i,j}}{D \times GIC_j}$$

2. Calculate the user profile U with regard to all GIC by multiplying the probabilities of all domains in the user's list (UD).

$$(2)\ U(GIC_j) = \prod_{i \in UD} P(D_i|GIC_j)$$

We did a 10-fold Cross-Validation to evaluate how successful interest profiles can be created based on this strategy. Since Google assigned interests binary but we calculated them based on probabilities we took the 8 most likely GIC to compare them with the GIC assigned by Google. Doing so an average of 5.5 of 8 GIC were assigned correctly. While this is not perfect it is good enough to build the dummy generation strategies on this since we do not need the most likely but the most unlikely GIC for the dummy generation strategy.

We therefore ordered the interests in the self-calculated profile and sorted them in reverse order leading to the least likely GIC as a basis for the DGS. To then select which websites should be used to obfuscate the profile we tested three strategies.

### B. RandomDGS

The RandomDGS is a simple dummy generation strategy that randomly selects URLs from a list and uses them as dummy traffic. The OpenDirectoryProject[8] was used as the dataset for RandomDGS. This set includes more than 4 million URLs which a community sorted into an ontology.

### C. StaticDGS

We created a list of website/interest combinations by surfing just to one website and afterwards checked whether an interest was assigned or not (direct relation). This list was used to select 25

---

[8] The Open Directory Project (http://www.dmoz.org/).

websites, 5 for each of the 5 least likely GIC to be added to the original URL-list.

*D. WeightedDGS*

For the weighted dummy generation strategy, we took into account the observed interest profiles and used the probabilities for a relation between a website and an interest as described in step 1 (see above) to also select 25 URLs for the 5 least likely GIC. Here indirect relations between websites and interests are also considered when selecting URLs for obfuscation

To test the effects of the DGS we used SYSTEM-A again and altered the users' domain lists by adding 25 URLs generated by one of the DGS. The DGS URLs and the Reddit URLs were added alternately between the 80$^{th}$ and 100$^{th}$ URLs resulting in a total of around 120 URLs. We then calculated the breadth of the resulting interest profiles as described above. The most successful of these DGS (staticDGS) was then tested again with 5 URLs for the 10 least likely GIC.

Because of the limitations described in IV.4 we also decided to re-run System-A without any changes, since 3 months had passed between the initial data collection described in section III and the tests of the SYSTEM-B.

*E. Results*

We found that using a DGS that learns from the combination of interests and websites is more effective than just adding any random website. Moreover, we proved that an observed relation of a website visit and an assigned interest (*staticDGS*) leads to more differences in the profile than not directly observed relations between a website and an interest (*weightedDGS*). The aim of obfuscation is that the interests based on the websites originally visited by a user are hidden in the interests produced by fake site visits. Table 3 shows details of different runs.

The effect of the obfuscation is weakened by the fact that the profiles have changed significantly within 3 months without any obfuscation strategy at all. The second row ("recrawl") in Table 3 shows that, there is a difference between the interests assigned in the original crawl and the begging of the test of SYSTEM-B. Profiles for the users only had an overlap of 58%. Therefore, there is a large variance in the profiles when the same websites are visited multiple times.

All resulted tests of SYSTEM-B resulted in profiles with lower breadth but vary in how much the profile was altered. Adding random URLs only caused little more difference compared to underlying noise measured with the recrawl. A measurable effect on the profiles visible for the informed obfuscation strategies. The effect of adding 25 URLs that support the 5 least likely interests worked similarly for direct and indirect correlations between sites and interests. The data therefore supports our hypothesis. The weightedDGS causes more interests to be lost (1.78). The staticDGS increases the overall breadth of a profile (B=7.6). Since for all strategies the resulting profiles are still far from being fully, we did an additional test with more URLs to be added to the profile. The row labeled *StaticDGS (+75)* show the results of the staticDGS when not 25 but 50 URLs were added that were taken from the 10 least like GIC. Although the strategy in higher number of new GIC, the increase in obfuscation is rather small taking into account that in this case 50% of the websites visited are dummy traffic.

*Table 3: Comparison of the effects on GIC-profile after adding 25 and 50 URLs with various DGS. Table shows averages; standard deviation in brackets.*

|  | GIC | new | lost |
|---|---|---|---|
| Orig | 8.25 (3.53) | / | / |
| Recrawl | 7.38 (2.24) | 1.29 (1.72) | 1.84 (1.32) |
| RandomDGS (+25) | 7.41 (2.08) | 1.90 (1.85) | 1.45 (1.36) |
| StaticDGS (+25) | 7.60 (1.94) | 2.29 (1.78) | 1.65 (1.48) |
| WeightedDGS (+25) | 7.31 (2.13) | 2.11 (1.62) | 1.78 (1.58) |
| StaticDGS (+50) | 7,97 (1,77) | 2,67 (1,63) | 1,64 (1,41) |

VI. DISCUSSION

Our obfuscation strategies have measurably influenced the interest profile created by google. However, the effect of adding 50% dummy traffic to a web surfing history, even when it is targeting at an opposite profile, is not very effective compared to the fact that visiting the same sites again 3 months later already results in a heavily changed profile. This may be related to the way online advertisements are sold (see IV.4). As described above a huge amount of websites within our test set was related to news websites, since those pages 'age' it might be, that the value they have to advertisers decreases and therefore do not serve as much targeted advertisements or are considered less relevant for an interest profile after the three month that passed between our tests.

There also remain uncertainties related to the fact that the profiling algorithms of Google are nontransparent. We do not know if Google has a Dummy Classification Algorithm (DCA) or a Profile Filtering Algorithm (PFA) in place which could be used to mitigate obfuscation. What we do know is that during the time of testing they did not use DCA methods comparable to those suggested by Balsa et al. [23]. The way that the crawler surfed the websites (staying 60 seconds on each page, a modified browser and without any interaction) would be filtered out by a *query based analysis* since it is easily distinguishable from more 'real' user interaction. Additionally, the chosen URLs that explicitly support unlikely interests would be an easy target for a *profile based analysis*. Also because obfuscation tools are still primarily used academically, Google can reasonably refrain from modifying their targeting until the need arises. Instead, it seems like Google is rather "optimistic" about the profiles. It broadens the GIC on low confidence data, e.g. if only one site out of a hundred is connected to a specific interest. We see this in the higher numbers of "new" than "lost" GIC. This is behavior is reasonable for an advertisement company as it increases the diversity of people to whom targeted ads will be shown.

One could generally argue that there is no need for obfuscation of GICs given that Google is as transparent and unstable in their interest assignings. Individuals can add or remove interests via the settings page to create a profile that matches whatever he or she wants. However, we do neither know whether the personalization differentiates between self-stated interests and those that emerge out of internal data sets and nor is it clear how long this service will be available since this option was recently removed for anonymous users.

## A. Limitations and Future Work

Our obfuscation system worked with interests associated with domains (instead of pages) so websites with a broader focus (like news portals) were not used for obfuscation. An improvement would be to identify connections between the content of a web page and the interest assigned by Google. In addition, the way the profiles were created within several hours does not imitate a human browsing pattern. Further improvements on this issue could provide more insights into how Google "forgets" interests and whether recent websites visits or regular visits have a larger effect on the profile.

Since we used bookmark lists only from reddit to perform the profile analysis, the gathered data about the distribution of interests is not generalizable. Nevertheless, the information about the extent of profiles sees plausible for any number of internet users. In addition, the dummy generation strategies presented in this paper are independent and could be enhanced by extending the input data. Unfortunately, Google removed the possibility to review interest profiles of anonymous users during a recent update while the option is still available for users with a google account. A future analysis therefore has to put extra effort in creating those accounts.

A further application of the resulting Bayes network could be found in using it to find pages that are out of the users focus helping them to escape a possible filter bubble. The results of this study can be used to e.g. develop an end-user web browser plug-in that not only increases transparency and control over existing profiles but will also empower users to impersonate a profile, fostering information literacy [44]. We are also considering how other information like demographic statistics that are commonly used to define a target group can be made transparent and controlled. This includes data about gender, age, location or income which are also provided by tracking services. All of this can be fruitful for "revitalizing serendipity" and offering users content that is normally hidden or down-ranked due to their browsing behavior.

## VII. SUMMARY

In this paper we argued that tracking and profiling of users is an elementary part of many web services. Profiles are created and used in a feedback loop were user actions change the user's profile, which changes the way she can interact with a service. Since there is a lack of options to opt-out of profiling there is a need for new privacy enhancing technologies that allow users to get transparency about and influence on their profiles. As part of this research we analyzed Google as one of the mayor players in web tracking and profiling business.

We studied the extent of tracking for 506 user profiles extracted from Reddit and found out that about 8 basic interests are assigned to each user by Google based on about 60 % of the web navigation trails the tracker was able to observe. This user-centered analysis is – to the knowledge of the author – new and has not been performed in a similar way. We then used this data to perform *informed obfuscation* by using different Dummy Generation Strategies (DGS) to influence a user's interest profile. We showed that, although there is an inherent noise in the profile data, these obfuscation strategies are more effective than adding random traffic.

Our study extends the knowledge about the practice of profiling users by a large online advertisement company. Gaining insights into these practices can increase the autonomy by counteracting the effects of services, which increasingly personalize their action without the knowledge or consent of users. The results can be used to increase transparency about tracking and profiling by offering information about what their profile looks like and which websites are related to which interest. In addition, the approach of informed obfuscation was proven to be more successful in supporting strategies to control profiling than randomly strategies that just try to hide within noisy data.

In future work we will evaluate how this data can be used to increase the awareness about the effects of online tracking by visualizing the profile. The promising results of our informed DGS are used to build an obfuscation browser extension to empower users to control their profiles in such a way as to be able change or extend them based on individual demands.